\documentclass[reprint,amsmath,amssymb,pra]{revtex4-2}

\usepackage{graphicx}
\usepackage{epstopdf}
\usepackage{dcolumn}
\usepackage{bm}
\usepackage{layouts}
\usepackage{ulem}
\usepackage{amsmath}
\usepackage{hyperref}
\hypersetup{
	colorlinks   = true, 
	urlcolor     = blue, 
	linkcolor    = blue, 
	citecolor   = blue 
}
\usepackage[dvipsnames]{xcolor}
\usepackage{float}

\def\ket#1{\left|#1\right\rangle}

\begin{document}

\title{Rapid generation and number-resolved detection of spinor Rubidium Bose-Einstein condensates}%


\author{Cebrail P\"ur$^1$}
\thanks{These authors contributed equally.}
\author{Mareike Hetzel$^{1}$} 
\thanks{These authors contributed equally.}
\author{} 
\email[Author to whom correspondence should be addressed. ]{hetzel@iqo.uni-hannover.de}
\author{Martin Quensen$^1$}
\author{Andreas H\"uper$^{1,2}$}
\author{Jiao Geng$^{3,4}$}
\author{Jens Kruse$^{1,2}$}
\author{Wolfgang Ertmer$^{1,2}$}
\author{Carsten Klempt$^{1,2}$}

\affiliation{$^1$Institut für Quantenoptik, Leibniz Universit\"at Hannover, Welfengarten 1, D-30167 Hannover, Germany  \\ $^2$Deutsches Zentrum f\"ur Luft- und Raumfahrt e.V. (DLR), Institut f\"ur Satellitengeod\"asie und Inertialsensorik (DLR-SI), Callinstraße 30b, D-30167 Hannover, Germany  \\ $^3$Key Laboratory of 3D Micro/Nano Fabrication and Characterization of Zhejiang Province, School of Engineering, Westlake University, 18 Shilongshan Road, Hangzhou 310024, Zhejiang Province, China \\ $^4$Institute of Advanced Technology, Westlake Institute for Advanced Study, 18 Shilongshan Road, Hangzhou 310024, Zhejiang Province, China }

\date{\today}

\begin{abstract}
High data acquisition rates and low-noise detection of ultracold neutral atoms present important challenges for the state tomography and interferometric application of entangled quantum states in  Bose-Einstein condensates.
In this article, we present a high-flux source of $^{87}$Rb Bose-Einstein condensates combined with a number-resolving detection. 
We create Bose-Einstein condensates of $2\times10^5$ atoms with no discernible thermal fraction within $3.3$~s using a hybrid evaporation approach in a magnetic/optical trap.
For the high-fidelity tomography of many-body quantum states in the spin degree of freedom~\cite{Hetzel2022}, it is desirable to select a single mode for a number-resolving detection.
We demonstrate the low-noise selection of subsamples of up to $16$ atoms and their subsequent detection with a counting noise below $0.2$ atoms.
The presented techniques offer an exciting path towards the creation and analysis of mesoscopic quantum states with unprecedented fidelities, and their exploitation for fundamental and metrological applications.
\end{abstract}

\maketitle

\section{\label{sec:level1}High-flux sources of Bose-Einstein condensates\protect}
Because of their well-controlled spatial mode and their phase coherence, Bose-Einstein condensates (BECs) of neutral atoms present a valuable resource for atom interferometry and quantum atom optics experiments in general.
Many of the applications, however, require short preparation times.
In atom interferometry, the preparation time defines the dead time of the sensor and therefore influences bandwidth and noise sensitivity.
Short preparation times are also important as they often dominate the data acquisition rate.
In our case, we wish to generate  entangled many-body states, and perform a high-fidelity state tomography.
The number of required measurements scales exponentially with the number of atoms.
Therefore, an accurate characterization of quantum states with an increasing number of atoms requires the acquisition of large data sets, during which the environmental conditions have to remain stable.
An improvement of the measurement rate not only improves the quality of the results, but in fact constitutes a requirement for scaling up the number of atoms in the generation of high-fidelity quantum states.

\begin{figure}
	\includegraphics{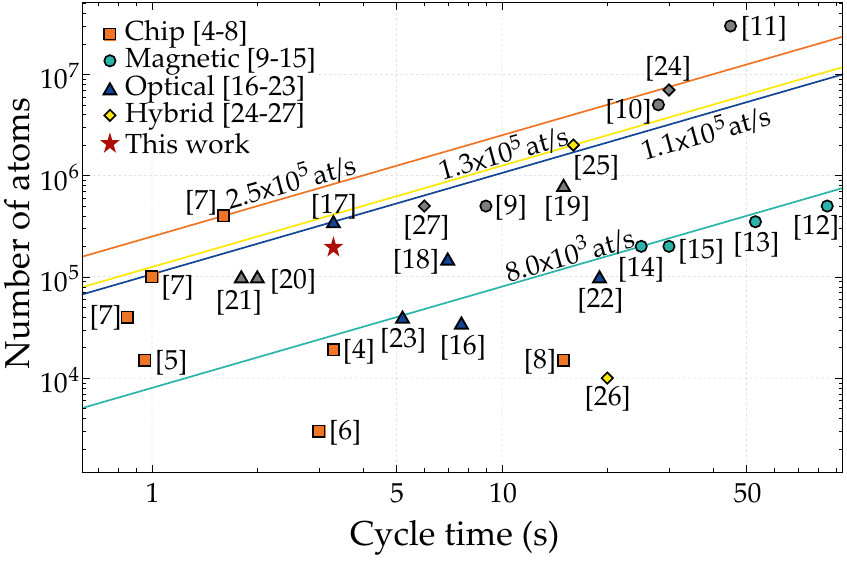}
	\caption{\label{fig:atomnumbervscycletime}
   Overview of the production time for different BEC sources with a given final atom number. 
   Orange squares are atom-chip based experiments, teal circles correspond to magnetically trapped ensembles, blue triangles show all-optical evaporation and yellow diamonds represent hybrid magnetic and optical methods. 
   Sources featuring other atomic species than rubidium are marked in grey. 
   The red star marks the result of this work corresponding to an atomic flux of $6 \times 10^4$ atoms per second. 
   Solid lines illustrate the atomic flux of the best performing experiments per category.}
\end{figure}

Besides few exemptions~\cite{Urvoy2019,Stellmer2013a}, BECs are typically realized by forced evaporative cooling in conservative potentials resulting from inhomogeneous magnetic or optical fields.
Figure~\ref{fig:atomnumbervscycletime} displays a selection of atom numbers and repetition rates obtained with atom-chip magnetic traps~ \cite{farkas2010a,farkas2014, horikoshi2006,Rudolph2015,abend2016}, macroscopic magnetic traps~\cite{Davis1995,Mewes1996,Naik2005,Esslinger1998,Kumar2015,Dubessy2012,Peil2003}, all-optical traps~\cite{barrett2001,kinoshita2005,clement2009,landini2012,Stellmer2013,roy2016,xie2018,condon2019}, and combinations of magnetic and optical traps~\cite{Colzi2018,Lin2009,Zaiser2011,Bouton2015}. 

For our experiments, we aim at BECs with $10^5$ rubidium atoms at a maximal repetition rate.
At the same time, the set-up is supposed to enable the selection and detection of subensembles with single-particle counting resolution. 
The inclusion of such a high-performance detection excludes the implementation of an atom chip, which would promise short preparation times, but leads to detrimental light scattering on the chip surface.

In this article, we present our experimental set-up which enables the generation of $2 \times 10^5$ rubidium atoms with a cycle time of $3.3$~s and the number-resolving detection of a subensemble in a specific spin state.
The BEC generation relies on a macroscopic quadrupole trap combined with a crossed-beam optical dipole trap (cODT).
The fluorescence detection is a fiber-based second generation from an earlier implementation~\cite{Hueper2021}.
This article describes the experimental set-up and the obtained results, also as a reference for the quantum detector tomography reported in the accompanying publication~\cite{Hetzel2022}.

This paper is organized as follows. 
In Section \ref{sec:level2}, we describe the details of our apparatus and our experimental sequence for fast BEC production.
Section \ref{sec:level3} presents the successive spin preparation of small ensembles by microwave transitions and subsequent optical removal of atoms. 
The number-resolving detection set-up is described in Section \ref{sec:level4}.
Section \ref{sec:level5} concludes with a summary and outlook.

\section{\label{sec:level2}Rapid creation of Bose-Einstein condensates}

\begin{figure*}[ht!]
	\includegraphics[width=1.0\linewidth]{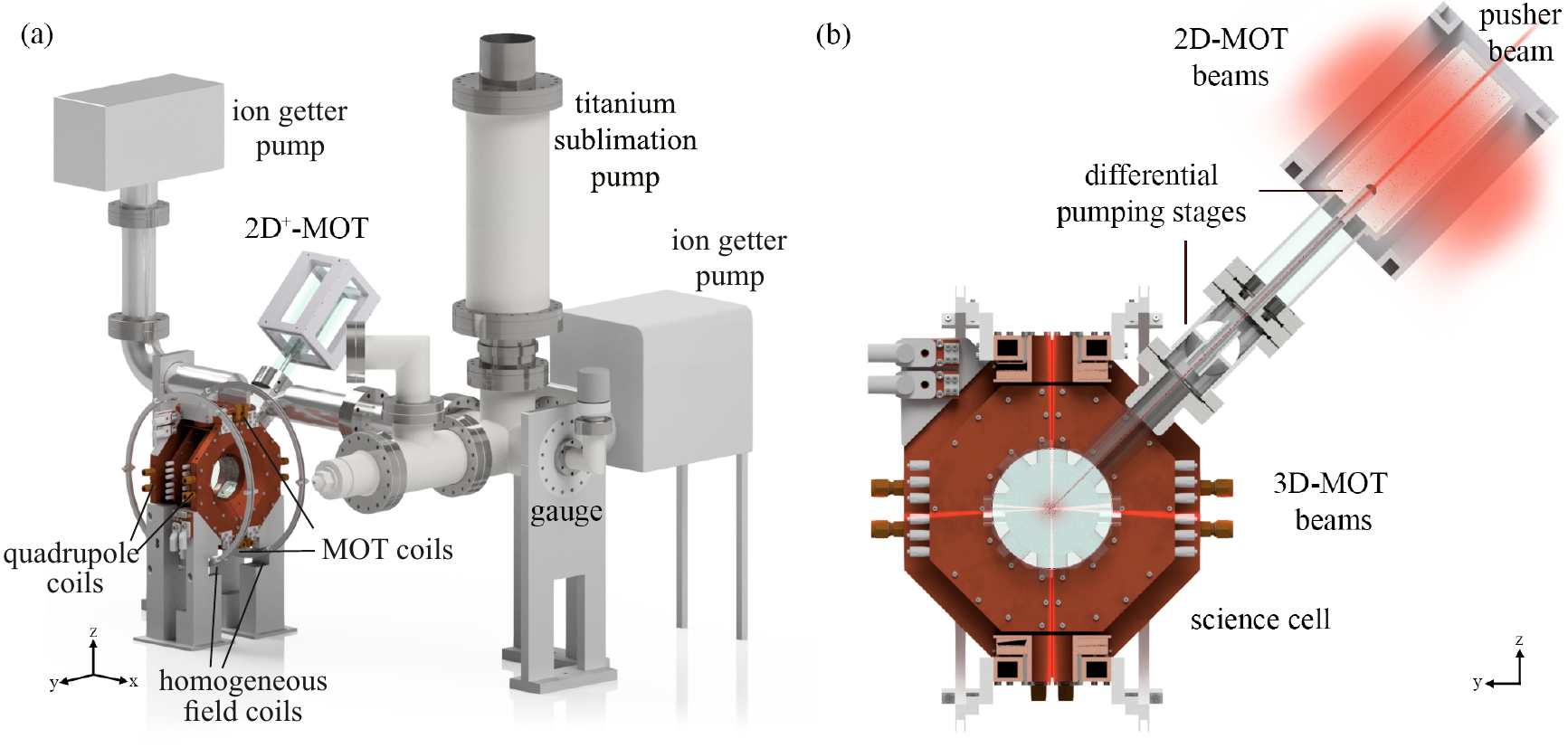}
	\caption{\label{fig:setup}(a) Computer-aided design of the ultra-high vacuum system with the two-chamber set-up and the coil system.
	The system contains two glass cells for the 2D$^+$-MOT and the 3D-MOT/BEC-generation, which are connected via two differential pumping stages at a $45^{\circ}$ angle.
	The connection tube and the science cell are individually pumped by two pump arrangements.
	The coil system surrounding the science cell is assembled by a pair of vertical 3D-MOT coils, a pair of horizontal quadrupole coils and a pair of orthogonally aligned Helmholtz coils.
	The science cell and the coil arrangement are optimized for a high-NA optical access along the x direction.
	(b) Cross-section of the two-chamber set-up and the coil system. Atoms are released from dispensers, building up the pressure inside the 2D$^+$-MOT. 
	A pusher beam guides the 2D$^+$-MOT beam through the two differential pumping stages into the science chamber, where they are captured and cooled.}
\end{figure*}

\begin{figure*}[ht!]
	\includegraphics{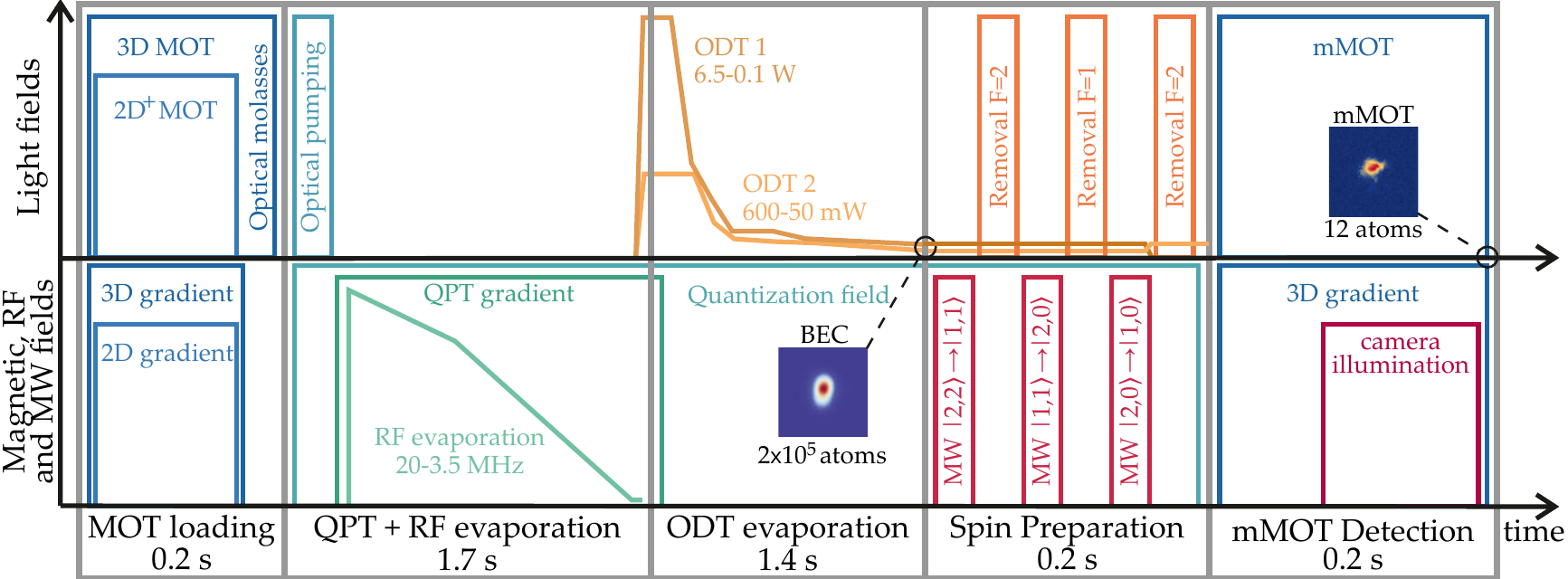}
	\caption{\label{fig:sequence}
	Schematic of the experimental sequence. 
	The upper and lower panels illustrate the light and magnetic fields, respectively. 
	The field strengths and the time axis are not to scale. 
	The BEC is created within $3.3$~s in a hybrid evaporation approach in a magnetic/optical trap. 
	Subsequently, the desired spin component is chosen and analyzed in the number-resolving mMOT set-up.  
	} 
\end{figure*}

The experimental set-up (Fig.~\ref{fig:setup}) consists of two glass cells which are connected by two sequential differential pumping stages.
The first, rectangular cell ($150\times60\times60$ mm) contains a two-dimensional magneto-optical trap (2D$^+$-MOT). 
The second, octagonal glass cell (seven $1$" and two $3$" viewports), contains a 3D-MOT. and further experiments are carried out.
The experimental sequence is initiated by loading the MOT and continues with the BEC creation and the mode-selective detection (Fig.~\ref{fig:sequence}).
The design of the 2D$^+$-MOT is described in Ref.~\cite{Jollenbeck2011}.
The 3D-MOT is operated with a coil pair of $20$ windings each in the vertical direction, yielding $12~$G/cm at $103$~A.
The cooling and detection laser light is generated by two external-cavity diode lasers~\cite{Baillard2006}, frequency-stabilized via modulation transfer spectroscopy~\cite{McCarron2008}, and 4 tapered amplifiers.
Optical fibers deliver the light to the vacuum set-up where the light power is actively stabilized.
Here six beams with a power of 35~mW and a waist of $7.5$~mm are formed and enter the glass cell via the 3-inch windows at an angle of $45^\circ$.
This beam configuration offers large optical access for detection with a numerical aperture $NA=0.42$.
The glass cell is anti-reflection-coated on both sides for 780nm-light to maximally suppress background light during the fluorescence detection.
The initial loading rate of our 3D-MOT is $2.4\times10^{10}$~atoms/s and within $200$~ms $4 \times 10^9$~atoms are captured.


After MOT loading, an optical molasses cools the atoms to $18~\mu K$.
The atoms are optically pumped to the level $\ket{F,m_F}=\ket{2,2}$ on the D2 $F=2 \rightarrow F'=2$ transition and loaded into a magnetic quadrupole trap (QPT).
The magnetic field gradient, generated by a coil pair with 62 windings each, is linearly ramped up from $58$ G/cm to $169$ G/cm in $50$ ms. 
The currents are measured by current transducers and actively stabilized to better than $10^{-4}$ relative stability.
The $1/e$ lifetime cannot be measured during the maximally allowed continuous operation of $10$~s and exceeds $150$~s. 
We perform evaporation by two linear radiofrequency (rf) ramps from $20$~MHz to $3.5$ MHz within $1.6$~s.

Afterwards, a crossed-beam optical dipole trapping potential is added with a trap center slightly below the magnetic trap center~\cite{Lin2009,KleineBuening2010}.
The optical dipole potential is generated by a $55$ W Coherent Mephisto MOPA laser with a wavelength of $1064$ nm. 
Two beams, whose intensity is each stabilized via an acousto-optic deflector (AOD) enter the glass cell via free-space optics along x and y-direction with waists of $70 \mu$m and $35 \mu$m, respectively. 
The power is increased to $6.5$ W and $600$ mW within $200$ ms, and the quadrupole field is ramped to zero within another $200$ ms.
Finally, the power of the dipole trap beams is decreased in six linear ramps to final values of $95$ mW and $45$ mW within $1.2$ s resulting in simulated trap frequencies of $(\omega_x,\omega_y,\omega_z,)=2\pi \times (60,160,150)~$Hz.



The evaporation is optimized for speed instead of atom number yielding an evaporation efficiency of $\gamma=1.7$.
After the final evaporation, we obtain BECs of $2\times10^5$ atoms with no discernible thermal fraction. 
The total BEC preparation takes 3.3~s.
Due to an insufficient cooling of the quadrupole trap coils, this cycle rate can so far only be maintained for 200 repetitions. 
An operation for longer times requires an additional 2~s cool-down time, or an improvement of the water cooling which is planned for the future.

\section{\label{sec:level3}Selection of a subensemble in a specific spin state}
After the rapid creation of a BEC, 
we aim at the generation of many-body spin states and their analysis by selecting one spin mode for number-resolved detection.
This spin-selective number detection is the prerequisite of the characterization of a coherent spin state in Ref.~\cite{Hetzel2022}.
Here, we demonstrate that a particular spin component can be chosen for detection with a fluorescence-based number counting.
The technique is demonstrated for up to 16~atoms, but can be extended to several hundred atoms in the near future.
While this number seems to be small at the first glance, we would like to highlight that this spin-selective detection allows for a tomography of multi-particle states with much larger atom numbers, as long as they are sufficiently polarized (e.g. spin-squeezed states).
As an example, we transfer a variable amount of atoms to the spin level $\ket{1,0}$ while we remove all remaining components in the $F=2$ manifold.
The atoms in level $\ket{1,0}$ are counted in a fiber-based miniature magneto-optical trap (mMOT).
The same technique can be implemented for a large variety of many-particle spin states to select a specific spin level for counting.
For the creation of many of these states, an ensemble in the spin level $\ket{1,0}$ serves as a starting point. 
Therefore, we characterize the necessary techniques by preparing and detecting a small fraction of the BEC in $\ket{1,0}$.
We select a subensemble of the BEC that we transfer into the spin level $\ket{1,1}$ by choosing the length of the MW pulse as a fraction of its $\pi$-pulse length. 
We remove the remaining components in the $F = 2$ manifold by a $100~\mu$s resonant cooling light push.
In the following, a MW $\pi$-pulse transfers the atoms to the spin level $\ket{2,0}$. 
To ensure there are no remaining fractions in F=1 caused by a non-perfect MW transfer, we employ a 100µs resonant repumping light push.
A final MW pulse transfers the atoms into the spin level $\ket{1,0}$ and another removal of atoms in $F=2$ ensures the detection of the desired components in $F=1$.
Between the optical removals short waiting times on the order of $10~$ms have been implemented to guarantee that the mechanical shutters open and close reliably.
The transfer of atoms to the level $\ket{1,0}$  from the initially prepared level $\ket{2,2}$ is carried out by a low-noise microwave source~\cite{Meyer2020}. 
The quantization axis is given by a magnetic field of $0.78$~G, which is actively stabilized to a magnetic field sensor, yielding a magnetic field noise of $170\;\mu$G.

A counting resolution on the single-atom level places strong requirements on the efficiency and selectivity of the final removal process. 
Two effects deteriorate the counting precision:
(i) Atoms expelled from the trap may collide with atoms that are selected for detection and remove them unintentionally.
(ii) The resonant light that is used to expel the $F=2$ atoms may pump atoms into the non-resonant $F=1$ levels before they leave the trap.
To avoid unwanted loss due to collisions during the optical removal, we transfer the atoms into a single-beam optical dipole trap. 
Pumping the atoms into a closed cycling transition reduces the probability of populating non-resonant states. 
Therefore, we use a homogeneous magnetic field of $6.7$~G and a $\sigma ^+$-polarized light beam for optical removal of the $F=2$ manifold before we detect the atoms in $F=1$ in our number-resolving mMOT set-up. 
To avoid recapturing the previously removed atoms, we implement a waiting time of $150$~ms after the final removal before detection. 

\section{\label{sec:level4}Number-resolving detection}
\begin{figure*}[ht!]
	\includegraphics{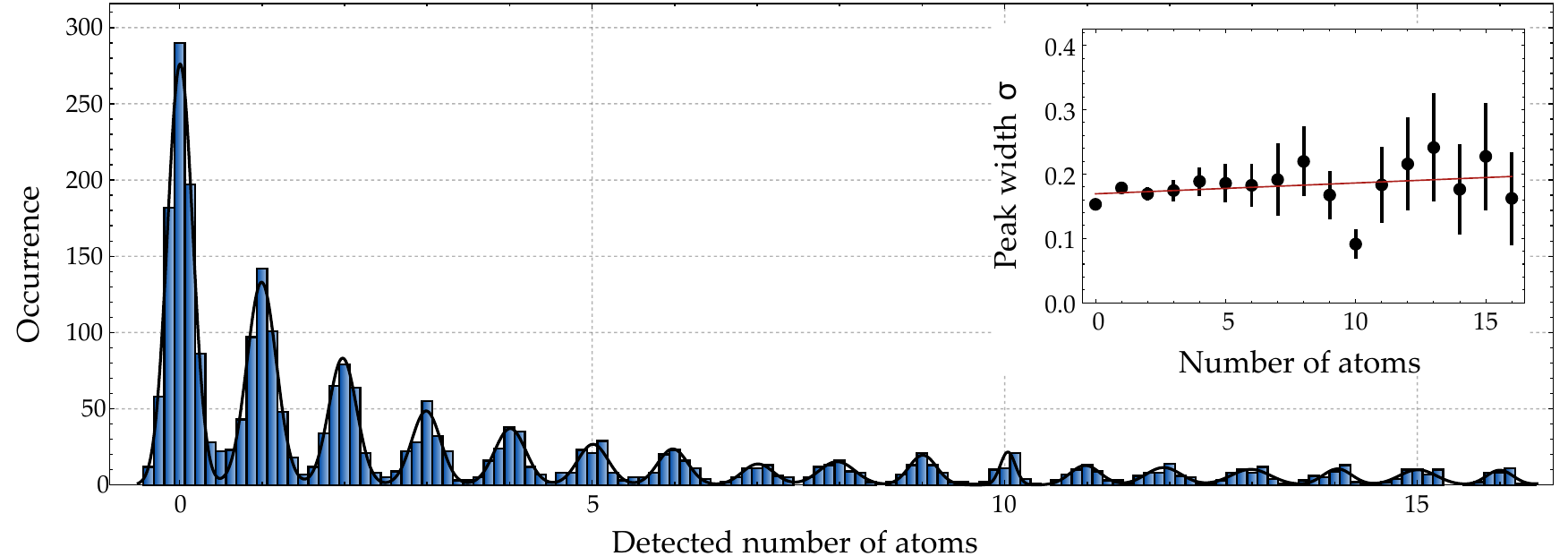}
	\caption{
	Histogram of the fluorescence signal for up to 16 atoms. 
    The signal (blue bars) shows a clear quantization for integer atom numbers.
    The individual peaks are fitted with Gaussian distributions (black line), whose widths are shown in the inset. 
    The linear fit (solid red line) is compatible with a  previously determined single-particle resolution limit of $400$ atoms.
	\label{fig:Histogram}} 
\end{figure*}

We detect the number of selected atoms in the $F=1$ manifold with an improved mMOT set-up that is based on the system described in Ref.~\cite{Hueper2021}.
The initial version of the detection set-up included beams distributed via free-space optics so that they could be superimposed with the larger MOT beams creating our 3D-MOT for BEC generation. 
This set-up proved to be capable of accurate atom counting but it showed long-term instabilities due to the long beam paths. 
The updated version therefore features six fiber-based beams minimizing the optical path lengths. 
Each fiber is individually mounted and its output intensity is actively stabilized, enabling a precise balancing of the intensities of the counter-propagating beams. 
The exact position of the mMOT depends on several parameters as the magnetic field minimum, the precise beam alignment and their intensity balance. 
To precisely overlap the mMOT with the BEC position, the mMOT beams are aligned onto the BEC position. 
Homogeneous magnetic fields in both horizontal directions shift the magnetic field minimum to the BEC position. 
The beam intensities and thereby the exact mMOT position have been optimized by a differential evolution algorithm~\cite{Geisel2013},
because the nontrivial interference of the beams leads to jumps of the mMOT position.
The mMOT is operated at a magnetic field gradient of $12$~G/cm with a detuning of $10$~MHz ($1.7 \Gamma$) and an intensity of $6$ times the saturation intensity.  
Due to spatial constraints, the new horizontal beams enter the science glass cell under an angle of $35^\circ$. 
This changed angle results in a higher background scattering in our detection objective and therefore in higher background noise $\sigma_{bg,new}=0.15$ compared to $0.03$ in the previous set-up~\cite{Hueper2021}. 
Fig.~\ref{fig:Histogram} shows a histogram of the recorded fluorescence signal for $2433$ repeated number measurements of atoms in the level $\ket{1,0}$. 
The histogram shows a clear quantization of the fluorescence signal, proving a counting resolution well below the single-atom limit.
The counting noise depends on the total number of atoms and is well described by Gaussian functions with a width $\sigma_N= 0.169 + 0.0017 N$.
Assuming a linear scaling, this allows for single-particle detection of more than 400~atoms, in correspondence to our earlier quantification~\cite{Hueper2021}.
The data can now be binned at half-integer binning limits, with a number assignment fidelity ranging from 99.7\% at 1 atom to 99.0\% at 15 atoms.

The mMOT has a finite probability to capture unwanted atoms from the background gas. 
These atoms are mainly caused by the frequent use of the 2D$^+$-MOT during the BEC creation, leading to a temporally increased background pressure. 
Atoms passing the small but still relevant trapping volume of the mMOT can be trapped and therefore counted in the detection process.
This results in a detection offset of $0.27$~atoms, which appears as a relevant signal in the quantum tomography of our number detection~\cite{Hetzel2022}.

\begin{figure*}[ht!]
	\includegraphics{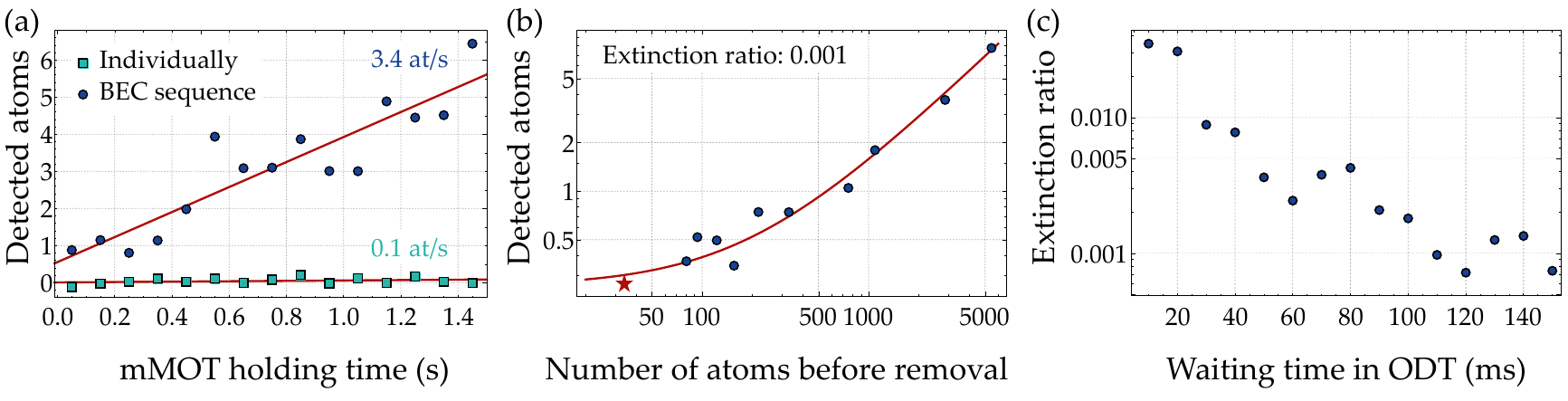}
	\caption{Characterization of the detection offset. (a) The frequent use of the 2D$^+$-MOT temporarily increases the background gas and therefore the mMOT loading rate from initially $0.1$ (cyan rectangles) to $3.4$~atoms/s (blue circles) after $7.5$~hours of operation. (b) The detection offset linearly depends on the number of atoms before removal.The red line shows a linear fit with an initial offset that is caused by a non-perfect optical removal of atoms. The red star marks the detection offset in Ref.~\cite{Hetzel2022}. (c) The extinction ratio of the optical removal improves with an increasing subsequent waiting time before detection. \label{fig:Push}} 
\end{figure*}

The effects contributing to the detection offset are characterized in Fig.~\ref{fig:Push}. 
Panel (a) shows the unwanted background loading of the mMOT, once in the beginning (cyan rectangles) and once after $7.5$~h of continuous operation (blue circles). 
The increase from $0.1$ to $3.4$ atoms per second extracted from linear fits is caused by the temporary increase of the background gas that builds up over time. 
We find that this build-up is generated by laser-cooled atoms in the 2D$^+$-MOT beam and not by thermal atoms passing the two differential pumping stages.

To maintain accurate atom number counting, the loading of atoms from the background gas has to be minimized which poses a limit to the total mMOT operation time. 
At the same time, our single-particle counting resolution demands long illumination times. 
A total mMOT operation time of $115~$ms including $65~$ms for illumination yields a reasonable compromise.
Fig.~\ref{fig:Push} (b) shows the linear dependence of the detection offset on the number of removed atoms $N_{rem}$. 
Non-perfect optical removals and increased recapturing of previously removed atoms both contribute to this increasing detection offset $N_{off}$ following $N_{off}=0.26+0.001 N_{rem}$. 
The red star illustrates the detection offset obtained in Ref.~\cite{Hetzel2022}. 
Fig.~\ref{fig:Push} (c) illustrates the extinction ratio in dependence of the waiting time after the optical removal. 
The waiting time of $150~$ms reduces the probability to recapture previously removed atoms in the mMOT by more than one order of magnitude.
In summary, the described detection enables a number-resolving tomography of quantum many-body states up to a few hundred atoms.

\section{\label{sec:level5}Conclusion \& Outlook}
In conclusion, we have presented a method for the generation and number-resolved detection of spinor BECs. 
We create BECs with $2\times10^5$ $^{87}$Rb atoms within $3.3$~s using a hybrid approach.
The high-flux atom source consists of a 2D$^+$-MOT combined with a 3D-MOT. 
After transfer into a magnetic quadrupole trap, a first rf-evaporation step is applied to increase the phase space density and to efficiently transfer the atoms into the optical dipole potential, in which fast efficient evaporation is performed.
Our experiment shows a high flux BEC creation of $6\times10^4$ atoms/s, which is close to the published record for Rb~\cite{Lin2009}, disregarding the atom chip experiments which do not provide sufficient optical access.
The cycle time can be further improved by decreasing the evaporation time in the crossed optical dipole trap. 
We can readily implement dynamically shaped potentials using our acousto-optical deflector set-up, which will give us independent control of the trap depth and trapping frequencies, leading to accelerated evaporation dynamics.

We select a single spin level for detection and optically remove residual atoms with an extinction ratio of 0.001. 
We resolve the created output states with a detection offset of 0.27~atoms and a number assignment fidelity of 99\% at 15 atoms.
The presented techniques pave the way for the high-fidelity tomography of polarized many-particle entangled states, such as single- and two-mode spin-squeezed states.

\begin{acknowledgments}
This work is supported by the QuantERA grants SQUEIS and MENTA.
We acknowledge financial support from the Deutsche Forschungsgemeinschaft (DFG, German Research Foundation) - Project-ID 274200144 - SFB 1227 DQ-mat within the project B01 and Germany’s Excellence Strategy - EXC-2123 QuantumFrontiers-Project-ID 390837967. 
M.Q. acknowledges support from the Hannover School for Nanotechnology (HSN).
\end{acknowledgments}

\bibliography{main,bib2}

\begin{thebibliography}{34}%
\makeatletter
\providecommand \@ifxundefined [1]{%
 \@ifx{#1\undefined}
}%
\providecommand \@ifnum [1]{%
 \ifnum #1\expandafter \@firstoftwo
 \else \expandafter \@secondoftwo
 \fi
}%
\providecommand \@ifx [1]{%
 \ifx #1\expandafter \@firstoftwo
 \else \expandafter \@secondoftwo
 \fi
}%
\providecommand \natexlab [1]{#1}%
\providecommand \enquote  [1]{``#1''}%
\providecommand \bibnamefont  [1]{#1}%
\providecommand \bibfnamefont [1]{#1}%
\providecommand \citenamefont [1]{#1}%
\providecommand \href@noop [0]{\@secondoftwo}%
\providecommand \href [0]{\begingroup \@sanitize@url \@href}%
\providecommand \@href[1]{\@@startlink{#1}\@@href}%
\providecommand \@@href[1]{\endgroup#1\@@endlink}%
\providecommand \@sanitize@url [0]{\catcode `\\12\catcode `\$12\catcode
  `\&12\catcode `\#12\catcode `\^12\catcode `\_12\catcode `\%12\relax}%
\providecommand \@@startlink[1]{}%
\providecommand \@@endlink[0]{}%
\providecommand \url  [0]{\begingroup\@sanitize@url \@url }%
\providecommand \@url [1]{\endgroup\@href {#1}{\urlprefix }}%
\providecommand \urlprefix  [0]{URL }%
\providecommand \Eprint [0]{\href }%
\providecommand \doibase [0]{https://doi.org/}%
\providecommand \selectlanguage [0]{\@gobble}%
\providecommand \bibinfo  [0]{\@secondoftwo}%
\providecommand \bibfield  [0]{\@secondoftwo}%
\providecommand \translation [1]{[#1]}%
\providecommand \BibitemOpen [0]{}%
\providecommand \bibitemStop [0]{}%
\providecommand \bibitemNoStop [0]{.\EOS\space}%
\providecommand \EOS [0]{\spacefactor3000\relax}%
\providecommand \BibitemShut  [1]{\csname bibitem#1\endcsname}%
\let\auto@bib@innerbib\@empty
\bibitem [{\citenamefont {Hetzel}\ \emph {et~al.}(2022)\citenamefont {Hetzel},
  \citenamefont {Pezz\`e}, \citenamefont {P\"ur}, \citenamefont {Quensen},
  \citenamefont {H\"uper}, \citenamefont {Geng}, \citenamefont {Kruse},
  \citenamefont {Santos}, \citenamefont {Ertmer}, \citenamefont {Smerzi},\ and\
  \citenamefont {Klempt}}]{Hetzel2022}%
  \BibitemOpen
  \bibfield  {author} {\bibinfo {author} {\bibfnamefont {M.}~\bibnamefont
  {Hetzel}}, \bibinfo {author} {\bibfnamefont {L.}~\bibnamefont {Pezz\`e}},
  \bibinfo {author} {\bibfnamefont {C.}~\bibnamefont {P\"ur}}, \bibinfo
  {author} {\bibfnamefont {M.}~\bibnamefont {Quensen}}, \bibinfo {author}
  {\bibfnamefont {A.}~\bibnamefont {H\"uper}}, \bibinfo {author} {\bibfnamefont
  {J.}~\bibnamefont {Geng}}, \bibinfo {author} {\bibfnamefont {J.}~\bibnamefont
  {Kruse}}, \bibinfo {author} {\bibfnamefont {L.}~\bibnamefont {Santos}},
  \bibinfo {author} {\bibfnamefont {W.}~\bibnamefont {Ertmer}}, \bibinfo
  {author} {\bibfnamefont {A.}~\bibnamefont {Smerzi}},\ and\ \bibinfo {author}
  {\bibfnamefont {C.}~\bibnamefont {Klempt}},\ }\bibfield  {title} {\bibinfo
  {title} {Tomography of a number-resolving detector by reconstruction of an
  atomic many-body quantum state},\ }\href {https://arxiv.org/abs/2207.01270}
  {\bibfield  {journal} {\bibinfo  {journal} {arXiv:2207.01270}\ } (\bibinfo
  {year} {2022})}\BibitemShut {NoStop}%
\bibitem [{\citenamefont {Urvoy}\ \emph {et~al.}(2019)\citenamefont {Urvoy},
  \citenamefont {Vendeiro}, \citenamefont {Ramette}, \citenamefont
  {Adiyatullin},\ and\ \citenamefont {Vuleti\ifmmode~\acute{c}\else
  \'{c}\fi{}}}]{Urvoy2019}%
  \BibitemOpen
  \bibfield  {author} {\bibinfo {author} {\bibfnamefont {A.}~\bibnamefont
  {Urvoy}}, \bibinfo {author} {\bibfnamefont {Z.}~\bibnamefont {Vendeiro}},
  \bibinfo {author} {\bibfnamefont {J.}~\bibnamefont {Ramette}}, \bibinfo
  {author} {\bibfnamefont {A.}~\bibnamefont {Adiyatullin}},\ and\ \bibinfo
  {author} {\bibfnamefont {V.}~\bibnamefont {Vuleti\ifmmode~\acute{c}\else
  \'{c}\fi{}}},\ }\bibfield  {title} {\bibinfo {title} {Direct laser cooling to
  {B}ose-{E}instein condensation in a dipole trap},\ }\href
  {https://doi.org/10.1103/PhysRevLett.122.203202} {\bibfield  {journal}
  {\bibinfo  {journal} {Phys. Rev. Lett.}\ }\textbf {\bibinfo {volume} {122}},\
  \bibinfo {pages} {203202} (\bibinfo {year} {2019})}\BibitemShut {NoStop}%
\bibitem [{\citenamefont {Stellmer}\ \emph
  {et~al.}(2013{\natexlab{a}})\citenamefont {Stellmer}, \citenamefont
  {Pasquiou}, \citenamefont {Grimm},\ and\ \citenamefont
  {Schreck}}]{Stellmer2013a}%
  \BibitemOpen
  \bibfield  {author} {\bibinfo {author} {\bibfnamefont {S.}~\bibnamefont
  {Stellmer}}, \bibinfo {author} {\bibfnamefont {B.}~\bibnamefont {Pasquiou}},
  \bibinfo {author} {\bibfnamefont {R.}~\bibnamefont {Grimm}},\ and\ \bibinfo
  {author} {\bibfnamefont {F.}~\bibnamefont {Schreck}},\ }\bibfield  {title}
  {\bibinfo {title} {Laser cooling to quantum degeneracy},\ }\href
  {https://doi.org/10.1103/PhysRevLett.110.263003} {\bibfield  {journal}
  {\bibinfo  {journal} {Phys. Rev. Lett.}\ }\textbf {\bibinfo {volume} {110}},\
  \bibinfo {pages} {263003} (\bibinfo {year} {2013}{\natexlab{a}})}\BibitemShut
  {NoStop}%
\bibitem [{\citenamefont {Farkas}\ \emph {et~al.}(2010)\citenamefont {Farkas},
  \citenamefont {Hudek}, \citenamefont {Salim}, \citenamefont {Segal},
  \citenamefont {Squires},\ and\ \citenamefont {Anderson}}]{farkas2010a}%
  \BibitemOpen
  \bibfield  {author} {\bibinfo {author} {\bibfnamefont {D.~M.}\ \bibnamefont
  {Farkas}}, \bibinfo {author} {\bibfnamefont {K.~M.}\ \bibnamefont {Hudek}},
  \bibinfo {author} {\bibfnamefont {E.~A.}\ \bibnamefont {Salim}}, \bibinfo
  {author} {\bibfnamefont {S.~R.}\ \bibnamefont {Segal}}, \bibinfo {author}
  {\bibfnamefont {M.~B.}\ \bibnamefont {Squires}},\ and\ \bibinfo {author}
  {\bibfnamefont {D.~Z.}\ \bibnamefont {Anderson}},\ }\bibfield  {title}
  {\bibinfo {title} {A compact, transportable, microchip-based system for high
  repetition rate production of {{Bose}}\textendash{{Einstein}} condensates},\
  }\href {https://doi.org/10.1063/1.3327812} {\bibfield  {journal} {\bibinfo
  {journal} {Appl. Phys. Lett.}\ }\textbf {\bibinfo {volume} {96}},\ \bibinfo
  {pages} {093102} (\bibinfo {year} {2010})}\BibitemShut {NoStop}%
\bibitem [{\citenamefont {Farkas}\ \emph {et~al.}(2014)\citenamefont {Farkas},
  \citenamefont {Salim},\ and\ \citenamefont {Ramirez-Serrano}}]{farkas2014}%
  \BibitemOpen
  \bibfield  {author} {\bibinfo {author} {\bibfnamefont {D.~M.}\ \bibnamefont
  {Farkas}}, \bibinfo {author} {\bibfnamefont {E.~A.}\ \bibnamefont {Salim}},\
  and\ \bibinfo {author} {\bibfnamefont {J.}~\bibnamefont {Ramirez-Serrano}},\
  }\bibfield  {title} {\bibinfo {title} {Production of {Rubidium}
  {Bose}-{Einstein} {Condensates} at a 1 {Hz} {Rate}},\ }\href
  {http://arxiv.org/abs/1403.4641} {\bibfield  {journal} {\bibinfo  {journal}
  {arXiv:1403.4641}\ } (\bibinfo {year} {2014})}\BibitemShut {NoStop}%
\bibitem [{\citenamefont {Horikoshi}\ and\ \citenamefont
  {Nakagawa}(2006)}]{horikoshi2006}%
  \BibitemOpen
  \bibfield  {author} {\bibinfo {author} {\bibfnamefont {M.}~\bibnamefont
  {Horikoshi}}\ and\ \bibinfo {author} {\bibfnamefont {K.}~\bibnamefont
  {Nakagawa}},\ }\bibfield  {title} {\bibinfo {title} {Atom chip based fast
  production of {Bose}-{Einstein} condensate},\ }\href
  {https://doi.org/10.1007/s00340-005-2083-z} {\bibfield  {journal} {\bibinfo
  {journal} {Applied Physics B}\ }\textbf {\bibinfo {volume} {82}},\ \bibinfo
  {pages} {363} (\bibinfo {year} {2006})}\BibitemShut {NoStop}%
\bibitem [{\citenamefont {Rudolph}\ \emph {et~al.}(2015)\citenamefont
  {Rudolph}, \citenamefont {Herr}, \citenamefont {Grzeschik}, \citenamefont
  {Sternke}, \citenamefont {Grote}, \citenamefont {Popp}, \citenamefont
  {Becker}, \citenamefont {M{\"u}ntinga}, \citenamefont {Ahlers}, \citenamefont
  {Peters}, \citenamefont {L{\"a}mmerzahl}, \citenamefont {Sengstock},
  \citenamefont {Gaaloul}, \citenamefont {Ertmer},\ and\ \citenamefont
  {Rasel}}]{Rudolph2015}%
  \BibitemOpen
  \bibfield  {author} {\bibinfo {author} {\bibfnamefont {J.}~\bibnamefont
  {Rudolph}}, \bibinfo {author} {\bibfnamefont {W.}~\bibnamefont {Herr}},
  \bibinfo {author} {\bibfnamefont {C.}~\bibnamefont {Grzeschik}}, \bibinfo
  {author} {\bibfnamefont {T.}~\bibnamefont {Sternke}}, \bibinfo {author}
  {\bibfnamefont {A.}~\bibnamefont {Grote}}, \bibinfo {author} {\bibfnamefont
  {M.}~\bibnamefont {Popp}}, \bibinfo {author} {\bibfnamefont {D.}~\bibnamefont
  {Becker}}, \bibinfo {author} {\bibfnamefont {H.}~\bibnamefont
  {M{\"u}ntinga}}, \bibinfo {author} {\bibfnamefont {H.}~\bibnamefont
  {Ahlers}}, \bibinfo {author} {\bibfnamefont {A.}~\bibnamefont {Peters}},
  \bibinfo {author} {\bibfnamefont {C.}~\bibnamefont {L{\"a}mmerzahl}},
  \bibinfo {author} {\bibfnamefont {K.}~\bibnamefont {Sengstock}}, \bibinfo
  {author} {\bibfnamefont {N.}~\bibnamefont {Gaaloul}}, \bibinfo {author}
  {\bibfnamefont {W.}~\bibnamefont {Ertmer}},\ and\ \bibinfo {author}
  {\bibfnamefont {E.~M.}\ \bibnamefont {Rasel}},\ }\bibfield  {title} {\bibinfo
  {title} {A high-flux {BEC} source for mobile atom interferometers},\ }\href
  {https://doi.org/10.1088/1367-2630/17/6/065001} {\bibfield  {journal}
  {\bibinfo  {journal} {New J. Phys.}\ }\textbf {\bibinfo {volume} {17}},\
  \bibinfo {pages} {065001} (\bibinfo {year} {2015})}\BibitemShut {NoStop}%
\bibitem [{\citenamefont {Abend}\ \emph {et~al.}(2016)\citenamefont {Abend},
  \citenamefont {Gebbe}, \citenamefont {Gersemann}, \citenamefont {Ahlers},
  \citenamefont {M{\"u}ntinga}, \citenamefont {Giese}, \citenamefont {Gaaloul},
  \citenamefont {Schubert}, \citenamefont {L{\"a}mmerzahl}, \citenamefont
  {Ertmer}, \citenamefont {Schleich},\ and\ \citenamefont {Rasel}}]{abend2016}%
  \BibitemOpen
  \bibfield  {author} {\bibinfo {author} {\bibfnamefont {S.}~\bibnamefont
  {Abend}}, \bibinfo {author} {\bibfnamefont {M.}~\bibnamefont {Gebbe}},
  \bibinfo {author} {\bibfnamefont {M.}~\bibnamefont {Gersemann}}, \bibinfo
  {author} {\bibfnamefont {H.}~\bibnamefont {Ahlers}}, \bibinfo {author}
  {\bibfnamefont {H.}~\bibnamefont {M{\"u}ntinga}}, \bibinfo {author}
  {\bibfnamefont {E.}~\bibnamefont {Giese}}, \bibinfo {author} {\bibfnamefont
  {N.}~\bibnamefont {Gaaloul}}, \bibinfo {author} {\bibfnamefont
  {C.}~\bibnamefont {Schubert}}, \bibinfo {author} {\bibfnamefont
  {C.}~\bibnamefont {L{\"a}mmerzahl}}, \bibinfo {author} {\bibfnamefont
  {W.}~\bibnamefont {Ertmer}}, \bibinfo {author} {\bibfnamefont {W.~P.}\
  \bibnamefont {Schleich}},\ and\ \bibinfo {author} {\bibfnamefont {E.~M.}\
  \bibnamefont {Rasel}},\ }\bibfield  {title} {\bibinfo {title} {Atom-chip
  fountain gravimeter},\ }\href
  {https://doi.org/10.1103/physrevlett.117.203003} {\bibfield  {journal}
  {\bibinfo  {journal} {Phys. Rev. Lett.}\ }\textbf {\bibinfo {volume} {117}},\
  \bibinfo {pages} {203003} (\bibinfo {year} {2016})}\BibitemShut {NoStop}%
\bibitem [{\citenamefont {Davis}\ \emph {et~al.}(1995)\citenamefont {Davis},
  \citenamefont {Mewes}, \citenamefont {Andrews}, \citenamefont {van Druten},
  \citenamefont {Durfee}, \citenamefont {Kurn},\ and\ \citenamefont
  {Ketterle}}]{Davis1995}%
  \BibitemOpen
  \bibfield  {author} {\bibinfo {author} {\bibfnamefont {K.~B.}\ \bibnamefont
  {Davis}}, \bibinfo {author} {\bibfnamefont {M.~O.}\ \bibnamefont {Mewes}},
  \bibinfo {author} {\bibfnamefont {M.~R.}\ \bibnamefont {Andrews}}, \bibinfo
  {author} {\bibfnamefont {N.~J.}\ \bibnamefont {van Druten}}, \bibinfo
  {author} {\bibfnamefont {D.~S.}\ \bibnamefont {Durfee}}, \bibinfo {author}
  {\bibfnamefont {D.~M.}\ \bibnamefont {Kurn}},\ and\ \bibinfo {author}
  {\bibfnamefont {W.}~\bibnamefont {Ketterle}},\ }\bibfield  {title} {\bibinfo
  {title} {{Bose-Einstein} condensation in a gas of sodium atoms},\ }\href
  {https://doi.org/10.1103/PhysRevLett.75.3969} {\bibfield  {journal} {\bibinfo
   {journal} {Phys. Rev. Lett.}\ }\textbf {\bibinfo {volume} {75}},\ \bibinfo
  {pages} {3969} (\bibinfo {year} {1995})}\BibitemShut {NoStop}%
\bibitem [{\citenamefont {Mewes}\ \emph {et~al.}(1996)\citenamefont {Mewes},
  \citenamefont {Andrews}, \citenamefont {van Druten}, \citenamefont {Kurn},
  \citenamefont {Durfee},\ and\ \citenamefont {Ketterle}}]{Mewes1996}%
  \BibitemOpen
  \bibfield  {author} {\bibinfo {author} {\bibfnamefont {M.-O.}\ \bibnamefont
  {Mewes}}, \bibinfo {author} {\bibfnamefont {M.~R.}\ \bibnamefont {Andrews}},
  \bibinfo {author} {\bibfnamefont {N.~J.}\ \bibnamefont {van Druten}},
  \bibinfo {author} {\bibfnamefont {D.~M.}\ \bibnamefont {Kurn}}, \bibinfo
  {author} {\bibfnamefont {D.~S.}\ \bibnamefont {Durfee}},\ and\ \bibinfo
  {author} {\bibfnamefont {W.}~\bibnamefont {Ketterle}},\ }\bibfield  {title}
  {\bibinfo {title} {{Bose-Einstein} condensation in a tightly confining dc
  magnetic trap},\ }\href {https://doi.org/10.1103/PhysRevLett.77.416}
  {\bibfield  {journal} {\bibinfo  {journal} {Phys. Rev. Lett.}\ }\textbf
  {\bibinfo {volume} {77}},\ \bibinfo {pages} {416} (\bibinfo {year}
  {1996})}\BibitemShut {NoStop}%
\bibitem [{\citenamefont {Naik}\ and\ \citenamefont {Raman}(2005)}]{Naik2005}%
  \BibitemOpen
  \bibfield  {author} {\bibinfo {author} {\bibfnamefont {D.~S.}\ \bibnamefont
  {Naik}}\ and\ \bibinfo {author} {\bibfnamefont {C.}~\bibnamefont {Raman}},\
  }\bibfield  {title} {\bibinfo {title} {Optically plugged quadrupole trap for
  {B}ose-{E}instein condensates},\ }\href
  {https://doi.org/10.1103/physreva.71.033617} {\bibfield  {journal} {\bibinfo
  {journal} {Phys. Rev. A}\ }\textbf {\bibinfo {volume} {71}},\ \bibinfo
  {pages} {033617} (\bibinfo {year} {2005})}\BibitemShut {NoStop}%
\bibitem [{\citenamefont {Esslinger}\ \emph {et~al.}(1998)\citenamefont
  {Esslinger}, \citenamefont {Bloch},\ and\ \citenamefont
  {H\"ansch}}]{Esslinger1998}%
  \BibitemOpen
  \bibfield  {author} {\bibinfo {author} {\bibfnamefont {T.}~\bibnamefont
  {Esslinger}}, \bibinfo {author} {\bibfnamefont {I.}~\bibnamefont {Bloch}},\
  and\ \bibinfo {author} {\bibfnamefont {T.~W.}\ \bibnamefont {H\"ansch}},\
  }\bibfield  {title} {\bibinfo {title} {{Bose-Einstein} condensation in a
  quadrupole-ioffe-configuration trap},\ }\href
  {https://doi.org/10.1103/PhysRevA.58.R2664} {\bibfield  {journal} {\bibinfo
  {journal} {Phys. Rev. A}\ }\textbf {\bibinfo {volume} {58}},\ \bibinfo
  {pages} {R2664} (\bibinfo {year} {1998})}\BibitemShut {NoStop}%
\bibitem [{\citenamefont {Kumar}\ \emph {et~al.}(2015)\citenamefont {Kumar},
  \citenamefont {Sarkar}, \citenamefont {Verma}, \citenamefont {Vishwakarma},
  \citenamefont {Noaman},\ and\ \citenamefont {Rapol}}]{Kumar2015}%
  \BibitemOpen
  \bibfield  {author} {\bibinfo {author} {\bibfnamefont {S.}~\bibnamefont
  {Kumar}}, \bibinfo {author} {\bibfnamefont {S.}~\bibnamefont {Sarkar}},
  \bibinfo {author} {\bibfnamefont {G.}~\bibnamefont {Verma}}, \bibinfo
  {author} {\bibfnamefont {C.}~\bibnamefont {Vishwakarma}}, \bibinfo {author}
  {\bibfnamefont {M.}~\bibnamefont {Noaman}},\ and\ \bibinfo {author}
  {\bibfnamefont {U.}~\bibnamefont {Rapol}},\ }\bibfield  {title} {\bibinfo
  {title} {Bose-einstein condensation in an electro-pneumatically transformed
  quadrupole-{I}offe magnetic trap},\ }\href
  {https://doi.org/10.1088/1367-2630/17/2/023062} {\bibfield  {journal}
  {\bibinfo  {journal} {New Journal of Physics}\ }\textbf {\bibinfo {volume}
  {17}},\ \bibinfo {pages} {023062} (\bibinfo {year} {2015})}\BibitemShut
  {NoStop}%
\bibitem [{\citenamefont {Dubessy}\ \emph {et~al.}(2012)\citenamefont
  {Dubessy}, \citenamefont {Merloti}, \citenamefont {Longchambon},
  \citenamefont {Pottie}, \citenamefont {Liennard}, \citenamefont {Perrin},
  \citenamefont {Lorent},\ and\ \citenamefont {Perrin}}]{Dubessy2012}%
  \BibitemOpen
  \bibfield  {author} {\bibinfo {author} {\bibfnamefont {R.}~\bibnamefont
  {Dubessy}}, \bibinfo {author} {\bibfnamefont {K.}~\bibnamefont {Merloti}},
  \bibinfo {author} {\bibfnamefont {L.}~\bibnamefont {Longchambon}}, \bibinfo
  {author} {\bibfnamefont {P.-E.}\ \bibnamefont {Pottie}}, \bibinfo {author}
  {\bibfnamefont {T.}~\bibnamefont {Liennard}}, \bibinfo {author}
  {\bibfnamefont {A.}~\bibnamefont {Perrin}}, \bibinfo {author} {\bibfnamefont
  {V.}~\bibnamefont {Lorent}},\ and\ \bibinfo {author} {\bibfnamefont
  {H.}~\bibnamefont {Perrin}},\ }\bibfield  {title} {\bibinfo {title}
  {Rubidium-87 {{Bose}}-{{Einstein}} condensate in an optically plugged
  quadrupole trap},\ }\bibfield  {journal} {\bibinfo  {journal} {Phys. Rev. A}\
  }\textbf {\bibinfo {volume} {85}},\ \href
  {https://doi.org/10.1103/PhysRevA.85.013643} {10.1103/PhysRevA.85.013643}
  (\bibinfo {year} {2012})\BibitemShut {NoStop}%
\bibitem [{\citenamefont {Peil}\ \emph {et~al.}(2003)\citenamefont {Peil},
  \citenamefont {Porto}, \citenamefont {Tolra}, \citenamefont {Obrecht},
  \citenamefont {King}, \citenamefont {Subbotin}, \citenamefont {Rolston},\
  and\ \citenamefont {Phillips}}]{Peil2003}%
  \BibitemOpen
  \bibfield  {author} {\bibinfo {author} {\bibfnamefont {S.}~\bibnamefont
  {Peil}}, \bibinfo {author} {\bibfnamefont {J.~V.}\ \bibnamefont {Porto}},
  \bibinfo {author} {\bibfnamefont {B.~L.}\ \bibnamefont {Tolra}}, \bibinfo
  {author} {\bibfnamefont {J.~M.}\ \bibnamefont {Obrecht}}, \bibinfo {author}
  {\bibfnamefont {B.~E.}\ \bibnamefont {King}}, \bibinfo {author}
  {\bibfnamefont {M.}~\bibnamefont {Subbotin}}, \bibinfo {author}
  {\bibfnamefont {S.~L.}\ \bibnamefont {Rolston}},\ and\ \bibinfo {author}
  {\bibfnamefont {W.~D.}\ \bibnamefont {Phillips}},\ }\bibfield  {title}
  {\bibinfo {title} {Patterned loading of a {B}ose-{E}instein condensate into
  an optical lattice},\ }\href {https://doi.org/10.1103/PhysRevA.67.051603}
  {\bibfield  {journal} {\bibinfo  {journal} {Phys. Rev. A}\ }\textbf {\bibinfo
  {volume} {67}},\ \bibinfo {pages} {051603} (\bibinfo {year}
  {2003})}\BibitemShut {NoStop}%
\bibitem [{\citenamefont {Barrett}\ \emph {et~al.}(2001)\citenamefont
  {Barrett}, \citenamefont {Sauer},\ and\ \citenamefont
  {Chapman}}]{barrett2001}%
  \BibitemOpen
  \bibfield  {author} {\bibinfo {author} {\bibfnamefont {M.~D.}\ \bibnamefont
  {Barrett}}, \bibinfo {author} {\bibfnamefont {J.~A.}\ \bibnamefont {Sauer}},\
  and\ \bibinfo {author} {\bibfnamefont {M.~S.}\ \bibnamefont {Chapman}},\
  }\bibfield  {title} {\bibinfo {title} {{All-Optical Formation of an Atomic
  {Bose-Einstein} Condensate}},\ }\href
  {https://doi.org/10.1103/PhysRevLett.87.010404} {\bibfield  {journal}
  {\bibinfo  {journal} {Phys. Rev. Lett.}\ }\textbf {\bibinfo {volume} {87}},\
  \bibinfo {pages} {010404} (\bibinfo {year} {2001})}\BibitemShut {NoStop}%
\bibitem [{\citenamefont {Kinoshita}\ \emph {et~al.}(2005)\citenamefont
  {Kinoshita}, \citenamefont {Wenger},\ and\ \citenamefont
  {Weiss}}]{kinoshita2005}%
  \BibitemOpen
  \bibfield  {author} {\bibinfo {author} {\bibfnamefont {T.}~\bibnamefont
  {Kinoshita}}, \bibinfo {author} {\bibfnamefont {T.}~\bibnamefont {Wenger}},\
  and\ \bibinfo {author} {\bibfnamefont {D.~S.}\ \bibnamefont {Weiss}},\
  }\bibfield  {title} {\bibinfo {title} {All-optical {Bose-Einstein}
  condensation using a compressible crossed dipole trap},\ }\href
  {https://doi.org/10.1103/PhysRevA.71.011602} {\bibfield  {journal} {\bibinfo
  {journal} {Phys. Rev. A}\ }\textbf {\bibinfo {volume} {71}},\ \bibinfo
  {pages} {011602} (\bibinfo {year} {2005})}\BibitemShut {NoStop}%
\bibitem [{\citenamefont {Cl\'ement}\ \emph {et~al.}(2009)\citenamefont
  {Cl\'ement}, \citenamefont {Brantut}, \citenamefont {Robert-de
  Saint-Vincent}, \citenamefont {Nyman}, \citenamefont {Aspect}, \citenamefont
  {Bourdel},\ and\ \citenamefont {Bouyer}}]{clement2009}%
  \BibitemOpen
  \bibfield  {author} {\bibinfo {author} {\bibfnamefont {J.-F.}\ \bibnamefont
  {Cl\'ement}}, \bibinfo {author} {\bibfnamefont {J.-P.}\ \bibnamefont
  {Brantut}}, \bibinfo {author} {\bibfnamefont {M.}~\bibnamefont {Robert-de
  Saint-Vincent}}, \bibinfo {author} {\bibfnamefont {R.~A.}\ \bibnamefont
  {Nyman}}, \bibinfo {author} {\bibfnamefont {A.}~\bibnamefont {Aspect}},
  \bibinfo {author} {\bibfnamefont {T.}~\bibnamefont {Bourdel}},\ and\ \bibinfo
  {author} {\bibfnamefont {P.}~\bibnamefont {Bouyer}},\ }\bibfield  {title}
  {\bibinfo {title} {{All-optical runaway evaporation to {Bose-Einstein}
  condensation}},\ }\href {https://doi.org/10.1103/PhysRevA.79.061406}
  {\bibfield  {journal} {\bibinfo  {journal} {Phys. Rev. A}\ }\textbf {\bibinfo
  {volume} {79}},\ \bibinfo {pages} {061406} (\bibinfo {year}
  {2009})}\BibitemShut {NoStop}%
\bibitem [{\citenamefont {Landini}\ \emph {et~al.}(2012)\citenamefont
  {Landini}, \citenamefont {Roy}, \citenamefont {Roati}, \citenamefont
  {Simoni}, \citenamefont {Inguscio}, \citenamefont {Modugno},\ and\
  \citenamefont {Fattori}}]{landini2012}%
  \BibitemOpen
  \bibfield  {author} {\bibinfo {author} {\bibfnamefont {M.}~\bibnamefont
  {Landini}}, \bibinfo {author} {\bibfnamefont {S.}~\bibnamefont {Roy}},
  \bibinfo {author} {\bibfnamefont {G.}~\bibnamefont {Roati}}, \bibinfo
  {author} {\bibfnamefont {A.}~\bibnamefont {Simoni}}, \bibinfo {author}
  {\bibfnamefont {M.}~\bibnamefont {Inguscio}}, \bibinfo {author}
  {\bibfnamefont {G.}~\bibnamefont {Modugno}},\ and\ \bibinfo {author}
  {\bibfnamefont {M.}~\bibnamefont {Fattori}},\ }\bibfield  {title} {\bibinfo
  {title} {Direct evaporative cooling of $^{39}${K} atoms to {Bose}-{Einstein}
  condensation},\ }\href {https://doi.org/10.1103/PhysRevA.86.033421}
  {\bibfield  {journal} {\bibinfo  {journal} {Physical Review A}\ }\textbf
  {\bibinfo {volume} {86}},\ \bibinfo {pages} {033421} (\bibinfo {year}
  {2012})}\BibitemShut {NoStop}%
\bibitem [{\citenamefont {Stellmer}\ \emph
  {et~al.}(2013{\natexlab{b}})\citenamefont {Stellmer}, \citenamefont {Grimm},\
  and\ \citenamefont {Schreck}}]{Stellmer2013}%
  \BibitemOpen
  \bibfield  {author} {\bibinfo {author} {\bibfnamefont {S.}~\bibnamefont
  {Stellmer}}, \bibinfo {author} {\bibfnamefont {R.}~\bibnamefont {Grimm}},\
  and\ \bibinfo {author} {\bibfnamefont {F.}~\bibnamefont {Schreck}},\
  }\bibfield  {title} {\bibinfo {title} {Production of quantum-degenerate
  strontium gases},\ }\href
  {https://link.aps.org/doi/10.1103/PhysRevA.87.013611} {\bibfield  {journal}
  {\bibinfo  {journal} {Physical Review A}\ }\textbf {\bibinfo {volume} {87}},\
  \bibinfo {pages} {013611} (\bibinfo {year} {2013}{\natexlab{b}})}\BibitemShut
  {NoStop}%
\bibitem [{\citenamefont {Roy}\ \emph {et~al.}(2016)\citenamefont {Roy},
  \citenamefont {Green}, \citenamefont {Bowler},\ and\ \citenamefont
  {Gupta}}]{roy2016}%
  \BibitemOpen
  \bibfield  {author} {\bibinfo {author} {\bibfnamefont {R.}~\bibnamefont
  {Roy}}, \bibinfo {author} {\bibfnamefont {A.}~\bibnamefont {Green}}, \bibinfo
  {author} {\bibfnamefont {R.}~\bibnamefont {Bowler}},\ and\ \bibinfo {author}
  {\bibfnamefont {S.}~\bibnamefont {Gupta}},\ }\bibfield  {title} {\bibinfo
  {title} {Rapid cooling to quantum degeneracy in dynamically shaped atom
  traps},\ }\href {https://doi.org/10.1103/PhysRevA.93.043403} {\bibfield
  {journal} {\bibinfo  {journal} {Phys. Rev. A}\ }\textbf {\bibinfo {volume}
  {93}},\ \bibinfo {pages} {043403} (\bibinfo {year} {2016})}\BibitemShut
  {NoStop}%
\bibitem [{\citenamefont {Xie}\ \emph {et~al.}(2018)\citenamefont {Xie},
  \citenamefont {Wang}, \citenamefont {Gou}, \citenamefont {Bu},\ and\
  \citenamefont {Yan}}]{xie2018}%
  \BibitemOpen
  \bibfield  {author} {\bibinfo {author} {\bibfnamefont {D.}~\bibnamefont
  {Xie}}, \bibinfo {author} {\bibfnamefont {D.}~\bibnamefont {Wang}}, \bibinfo
  {author} {\bibfnamefont {W.}~\bibnamefont {Gou}}, \bibinfo {author}
  {\bibfnamefont {W.}~\bibnamefont {Bu}},\ and\ \bibinfo {author}
  {\bibfnamefont {B.}~\bibnamefont {Yan}},\ }\bibfield  {title} {\bibinfo
  {title} {Fast production of rubidium {Bose}-{Einstein} condensate in a dimple
  trap},\ }\href {https://doi.org/10.1364/JOSAB.35.000500} {\bibfield
  {journal} {\bibinfo  {journal} {JOSA B}\ }\textbf {\bibinfo {volume} {35}},\
  \bibinfo {pages} {500} (\bibinfo {year} {2018})}\BibitemShut {NoStop}%
\bibitem [{\citenamefont {Condon}\ \emph {et~al.}(2019)\citenamefont {Condon},
  \citenamefont {Rabault}, \citenamefont {Barrett}, \citenamefont {Chichet},
  \citenamefont {Arguel}, \citenamefont {Eneriz-Imaz}, \citenamefont {Naik},
  \citenamefont {Bertoldi}, \citenamefont {Battelier}, \citenamefont {Bouyer},\
  and\ \citenamefont {Landragin}}]{condon2019}%
  \BibitemOpen
  \bibfield  {author} {\bibinfo {author} {\bibfnamefont {G.}~\bibnamefont
  {Condon}}, \bibinfo {author} {\bibfnamefont {M.}~\bibnamefont {Rabault}},
  \bibinfo {author} {\bibfnamefont {B.}~\bibnamefont {Barrett}}, \bibinfo
  {author} {\bibfnamefont {L.}~\bibnamefont {Chichet}}, \bibinfo {author}
  {\bibfnamefont {R.}~\bibnamefont {Arguel}}, \bibinfo {author} {\bibfnamefont
  {H.}~\bibnamefont {Eneriz-Imaz}}, \bibinfo {author} {\bibfnamefont
  {D.}~\bibnamefont {Naik}}, \bibinfo {author} {\bibfnamefont {A.}~\bibnamefont
  {Bertoldi}}, \bibinfo {author} {\bibfnamefont {B.}~\bibnamefont {Battelier}},
  \bibinfo {author} {\bibfnamefont {P.}~\bibnamefont {Bouyer}},\ and\ \bibinfo
  {author} {\bibfnamefont {A.}~\bibnamefont {Landragin}},\ }\bibfield  {title}
  {\bibinfo {title} {All-{Optical} {Bose}-{Einstein} {Condensates} in
  {Microgravity}},\ }\href {https://doi.org/10.1103/PhysRevLett.123.240402}
  {\bibfield  {journal} {\bibinfo  {journal} {Physical Review Letters}\
  }\textbf {\bibinfo {volume} {123}},\ \bibinfo {pages} {240402} (\bibinfo
  {year} {2019})}\BibitemShut {NoStop}%
\bibitem [{\citenamefont {Colzi}\ \emph {et~al.}(2018)\citenamefont {Colzi},
  \citenamefont {Fava}, \citenamefont {Barbiero}, \citenamefont {Mordini},
  \citenamefont {Lamporesi},\ and\ \citenamefont {Ferrari}}]{Colzi2018}%
  \BibitemOpen
  \bibfield  {author} {\bibinfo {author} {\bibfnamefont {G.}~\bibnamefont
  {Colzi}}, \bibinfo {author} {\bibfnamefont {E.}~\bibnamefont {Fava}},
  \bibinfo {author} {\bibfnamefont {M.}~\bibnamefont {Barbiero}}, \bibinfo
  {author} {\bibfnamefont {C.}~\bibnamefont {Mordini}}, \bibinfo {author}
  {\bibfnamefont {G.}~\bibnamefont {Lamporesi}},\ and\ \bibinfo {author}
  {\bibfnamefont {G.}~\bibnamefont {Ferrari}},\ }\bibfield  {title} {\bibinfo
  {title} {Production of large {B}ose-{E}instein condensates in a
  magnetic-shield-compatible hybrid trap},\ }\href
  {https://doi.org/10.1103/physreva.97.053625} {\bibfield  {journal} {\bibinfo
  {journal} {Physical Review A}\ }\textbf {\bibinfo {volume} {97}},\ \bibinfo
  {pages} {053625} (\bibinfo {year} {2018})}\BibitemShut {NoStop}%
\bibitem [{\citenamefont {Lin}\ \emph {et~al.}(2009)\citenamefont {Lin},
  \citenamefont {Perry}, \citenamefont {Compton}, \citenamefont {Spielman},\
  and\ \citenamefont {Porto}}]{Lin2009}%
  \BibitemOpen
  \bibfield  {author} {\bibinfo {author} {\bibfnamefont {Y.-J.}\ \bibnamefont
  {Lin}}, \bibinfo {author} {\bibfnamefont {A.~R.}\ \bibnamefont {Perry}},
  \bibinfo {author} {\bibfnamefont {R.~L.}\ \bibnamefont {Compton}}, \bibinfo
  {author} {\bibfnamefont {I.~B.}\ \bibnamefont {Spielman}},\ and\ \bibinfo
  {author} {\bibfnamefont {J.~V.}\ \bibnamefont {Porto}},\ }\bibfield  {title}
  {\bibinfo {title} {Rapid production of $^{87}${R}b {Bose-Einstein}
  condensates in a combined magnetic and optical potential},\ }\href
  {https://doi.org/10.1103/PhysRevA.79.063631} {\bibfield  {journal} {\bibinfo
  {journal} {Phys. Rev. A}\ }\textbf {\bibinfo {volume} {79}},\ \bibinfo
  {pages} {063631} (\bibinfo {year} {2009})}\BibitemShut {NoStop}%
\bibitem [{\citenamefont {Zaiser}\ \emph {et~al.}(2011)\citenamefont {Zaiser},
  \citenamefont {Hartwig}, \citenamefont {Schlippert}, \citenamefont {Velte},
  \citenamefont {Winter}, \citenamefont {Lebedev}, \citenamefont {Ertmer},\
  and\ \citenamefont {Rasel}}]{Zaiser2011}%
  \BibitemOpen
  \bibfield  {author} {\bibinfo {author} {\bibfnamefont {M.}~\bibnamefont
  {Zaiser}}, \bibinfo {author} {\bibfnamefont {J.}~\bibnamefont {Hartwig}},
  \bibinfo {author} {\bibfnamefont {D.}~\bibnamefont {Schlippert}}, \bibinfo
  {author} {\bibfnamefont {U.}~\bibnamefont {Velte}}, \bibinfo {author}
  {\bibfnamefont {N.}~\bibnamefont {Winter}}, \bibinfo {author} {\bibfnamefont
  {V.}~\bibnamefont {Lebedev}}, \bibinfo {author} {\bibfnamefont
  {W.}~\bibnamefont {Ertmer}},\ and\ \bibinfo {author} {\bibfnamefont {E.~M.}\
  \bibnamefont {Rasel}},\ }\bibfield  {title} {\bibinfo {title} {Simple method
  for generating {Bose-Einstein} condensates in a weak hybrid trap},\ }\href
  {https://doi.org/10.1103/PhysRevA.83.035601} {\bibfield  {journal} {\bibinfo
  {journal} {Phys. Rev. A}\ }\textbf {\bibinfo {volume} {83}},\ \bibinfo
  {pages} {035601} (\bibinfo {year} {2011})}\BibitemShut {NoStop}%
\bibitem [{\citenamefont {Bouton}\ \emph {et~al.}(2015)\citenamefont {Bouton},
  \citenamefont {Chang}, \citenamefont {Hoendervanger}, \citenamefont
  {Nogrette}, \citenamefont {Aspect}, \citenamefont {Westbrook},\ and\
  \citenamefont {Cl\'ement}}]{Bouton2015}%
  \BibitemOpen
  \bibfield  {author} {\bibinfo {author} {\bibfnamefont {Q.}~\bibnamefont
  {Bouton}}, \bibinfo {author} {\bibfnamefont {R.}~\bibnamefont {Chang}},
  \bibinfo {author} {\bibfnamefont {A.~L.}\ \bibnamefont {Hoendervanger}},
  \bibinfo {author} {\bibfnamefont {F.}~\bibnamefont {Nogrette}}, \bibinfo
  {author} {\bibfnamefont {A.}~\bibnamefont {Aspect}}, \bibinfo {author}
  {\bibfnamefont {C.~I.}\ \bibnamefont {Westbrook}},\ and\ \bibinfo {author}
  {\bibfnamefont {D.}~\bibnamefont {Cl\'ement}},\ }\bibfield  {title} {\bibinfo
  {title} {Fast production of {Bose}-{Einstein} condensates of metastable
  helium},\ }\href {https://doi.org/10.1103/PhysRevA.91.061402} {\bibfield
  {journal} {\bibinfo  {journal} {Physical Review A}\ }\textbf {\bibinfo
  {volume} {91}},\ \bibinfo {pages} {061402} (\bibinfo {year}
  {2015})}\BibitemShut {NoStop}%
\bibitem [{\citenamefont {H\"uper}\ \emph {et~al.}(2021)\citenamefont
  {H\"uper}, \citenamefont {P\"ur}, \citenamefont {Hetzel}, \citenamefont
  {Geng}, \citenamefont {Peise}, \citenamefont {Kruse}, \citenamefont
  {Kristensen}, \citenamefont {Ertmer}, \citenamefont {Arlt},\ and\
  \citenamefont {Klempt}}]{Hueper2021}%
  \BibitemOpen
  \bibfield  {author} {\bibinfo {author} {\bibfnamefont {A.}~\bibnamefont
  {H\"uper}}, \bibinfo {author} {\bibfnamefont {C.}~\bibnamefont {P\"ur}},
  \bibinfo {author} {\bibfnamefont {M.}~\bibnamefont {Hetzel}}, \bibinfo
  {author} {\bibfnamefont {J.}~\bibnamefont {Geng}}, \bibinfo {author}
  {\bibfnamefont {J.}~\bibnamefont {Peise}}, \bibinfo {author} {\bibfnamefont
  {I.}~\bibnamefont {Kruse}}, \bibinfo {author} {\bibfnamefont
  {M.}~\bibnamefont {Kristensen}}, \bibinfo {author} {\bibfnamefont
  {W.}~\bibnamefont {Ertmer}}, \bibinfo {author} {\bibfnamefont
  {J.}~\bibnamefont {Arlt}},\ and\ \bibinfo {author} {\bibfnamefont
  {C.}~\bibnamefont {Klempt}},\ }\bibfield  {title} {\bibinfo {title}
  {Number-resolved preparation of mesoscopic atomic ensembles},\ }\href
  {https://doi.org/10.1088/1367-2630/abd058} {\bibfield  {journal} {\bibinfo
  {journal} {New J. Phys.}\ }\textbf {\bibinfo {volume} {23}},\ \bibinfo
  {pages} {113046} (\bibinfo {year} {2021})}\BibitemShut {NoStop}%
\bibitem [{\citenamefont {J\"ollenbeck}\ \emph {et~al.}(2011)\citenamefont
  {J\"ollenbeck}, \citenamefont {Mahnke}, \citenamefont {Randoll},
  \citenamefont {Ertmer}, \citenamefont {Arlt},\ and\ \citenamefont
  {Klempt}}]{Jollenbeck2011}%
  \BibitemOpen
  \bibfield  {author} {\bibinfo {author} {\bibfnamefont {S.}~\bibnamefont
  {J\"ollenbeck}}, \bibinfo {author} {\bibfnamefont {J.}~\bibnamefont
  {Mahnke}}, \bibinfo {author} {\bibfnamefont {R.}~\bibnamefont {Randoll}},
  \bibinfo {author} {\bibfnamefont {W.}~\bibnamefont {Ertmer}}, \bibinfo
  {author} {\bibfnamefont {J.}~\bibnamefont {Arlt}},\ and\ \bibinfo {author}
  {\bibfnamefont {C.}~\bibnamefont {Klempt}},\ }\bibfield  {title} {\bibinfo
  {title} {Hexapole-compensated magneto-optical trap on a mesoscopic atom
  chip},\ }\href {https://doi.org/10.1103/PhysRevA.83.043406} {\bibfield
  {journal} {\bibinfo  {journal} {Phys. Rev. A}\ }\textbf {\bibinfo {volume}
  {83}},\ \bibinfo {pages} {043406} (\bibinfo {year} {2011})}\BibitemShut
  {NoStop}%
\bibitem [{\citenamefont {Baillard}\ \emph {et~al.}(2006)\citenamefont
  {Baillard}, \citenamefont {Gauguet}, \citenamefont {Bize}, \citenamefont
  {Lemonde}, \citenamefont {Laurent}, \citenamefont {Clairon},\ and\
  \citenamefont {Rosenbusch}}]{Baillard2006}%
  \BibitemOpen
  \bibfield  {author} {\bibinfo {author} {\bibfnamefont {X.}~\bibnamefont
  {Baillard}}, \bibinfo {author} {\bibfnamefont {A.}~\bibnamefont {Gauguet}},
  \bibinfo {author} {\bibfnamefont {S.}~\bibnamefont {Bize}}, \bibinfo {author}
  {\bibfnamefont {P.}~\bibnamefont {Lemonde}}, \bibinfo {author} {\bibfnamefont
  {P.}~\bibnamefont {Laurent}}, \bibinfo {author} {\bibfnamefont
  {A.}~\bibnamefont {Clairon}},\ and\ \bibinfo {author} {\bibfnamefont
  {P.}~\bibnamefont {Rosenbusch}},\ }\bibfield  {title} {\bibinfo {title}
  {Interference-filter-stabilized external-cavity diode lasers},\ }\href
  {https://doi.org/10.1016/j.optcom.2006.05.011} {\bibfield  {journal}
  {\bibinfo  {journal} {Opt. Comm.}\ }\textbf {\bibinfo {volume} {266}},\
  \bibinfo {pages} {609 } (\bibinfo {year} {2006})}\BibitemShut {NoStop}%
\bibitem [{\citenamefont {McCarron}\ \emph {et~al.}(2008)\citenamefont
  {McCarron}, \citenamefont {King},\ and\ \citenamefont
  {Cornish}}]{McCarron2008}%
  \BibitemOpen
  \bibfield  {author} {\bibinfo {author} {\bibfnamefont {D.~J.}\ \bibnamefont
  {McCarron}}, \bibinfo {author} {\bibfnamefont {S.~A.}\ \bibnamefont {King}},\
  and\ \bibinfo {author} {\bibfnamefont {S.~L.}\ \bibnamefont {Cornish}},\
  }\bibfield  {title} {\bibinfo {title} {Modulation transfer spectroscopy in
  atomic rubidium},\ }\href {https://doi.org/10.1088/0957-0233/19/10/105601}
  {\bibfield  {journal} {\bibinfo  {journal} {Measurement Science and
  Technology}\ }\textbf {\bibinfo {volume} {19}},\ \bibinfo {pages} {105601}
  (\bibinfo {year} {2008})}\BibitemShut {NoStop}%
\bibitem [{\citenamefont {Kleine~B\"u{}ning}\ \emph {et~al.}(2010)\citenamefont
  {Kleine~B\"u{}ning}, \citenamefont {Will}, \citenamefont {Ertmer},
  \citenamefont {Klempt},\ and\ \citenamefont {Arlt}}]{KleineBuening2010}%
  \BibitemOpen
  \bibfield  {author} {\bibinfo {author} {\bibfnamefont {G.}~\bibnamefont
  {Kleine~B\"u{}ning}}, \bibinfo {author} {\bibfnamefont {J.}~\bibnamefont
  {Will}}, \bibinfo {author} {\bibfnamefont {W.}~\bibnamefont {Ertmer}},
  \bibinfo {author} {\bibfnamefont {C.}~\bibnamefont {Klempt}},\ and\ \bibinfo
  {author} {\bibfnamefont {J.}~\bibnamefont {Arlt}},\ }\bibfield  {title}
  {\bibinfo {title} {A slow gravity compensated atom laser},\ }\href
  {https://doi.org/10.1007/s00340-010-4078-7} {\bibfield  {journal} {\bibinfo
  {journal} {Appl. Phys. B}\ }\textbf {\bibinfo {volume} {100}},\ \bibinfo
  {pages} {117} (\bibinfo {year} {2010})}\BibitemShut {NoStop}%
\bibitem [{\citenamefont {Meyer}\ \emph {et~al.}(2020)\citenamefont {Meyer},
  \citenamefont {Idel}, \citenamefont {Anders}, \citenamefont {Peise},\ and\
  \citenamefont {Klempt}}]{Meyer2020}%
  \BibitemOpen
  \bibfield  {author} {\bibinfo {author} {\bibfnamefont {B.}~\bibnamefont
  {Meyer}}, \bibinfo {author} {\bibfnamefont {A.}~\bibnamefont {Idel}},
  \bibinfo {author} {\bibfnamefont {F.}~\bibnamefont {Anders}}, \bibinfo
  {author} {\bibfnamefont {J.}~\bibnamefont {Peise}},\ and\ \bibinfo {author}
  {\bibfnamefont {C.}~\bibnamefont {Klempt}},\ }\bibfield  {title} {\bibinfo
  {title} {Dynamical low-noise microwave source for cold atom experiments},\
  }\href {https://arxiv.org/abs/2003.10989} {\bibfield  {journal} {\bibinfo
  {journal} {arXiv:2003.10989}\ } (\bibinfo {year} {2020})}\BibitemShut
  {NoStop}%
\bibitem [{\citenamefont {Geisel}\ \emph {et~al.}(2013)\citenamefont {Geisel},
  \citenamefont {Cordes}, \citenamefont {Mahnke}, \citenamefont {J\"ollenbeck},
  \citenamefont {Ostermann}, \citenamefont {Arlt}, \citenamefont {Ertmer},\
  and\ \citenamefont {Klempt}}]{Geisel2013}%
  \BibitemOpen
  \bibfield  {author} {\bibinfo {author} {\bibfnamefont {I.}~\bibnamefont
  {Geisel}}, \bibinfo {author} {\bibfnamefont {K.}~\bibnamefont {Cordes}},
  \bibinfo {author} {\bibfnamefont {J.}~\bibnamefont {Mahnke}}, \bibinfo
  {author} {\bibfnamefont {S.}~\bibnamefont {J\"ollenbeck}}, \bibinfo {author}
  {\bibfnamefont {J.}~\bibnamefont {Ostermann}}, \bibinfo {author}
  {\bibfnamefont {J.}~\bibnamefont {Arlt}}, \bibinfo {author} {\bibfnamefont
  {W.}~\bibnamefont {Ertmer}},\ and\ \bibinfo {author} {\bibfnamefont
  {C.}~\bibnamefont {Klempt}},\ }\bibfield  {title} {\bibinfo {title}
  {Evolutionary optimization of an experimental apparatus},\ }\href
  {https://doi.org/http://dx.doi.org/10.1063/1.4808213} {\bibfield  {journal}
  {\bibinfo  {journal} {Appl. Phys. Lett.}\ }\textbf {\bibinfo {volume}
  {102}},\  (\bibinfo {year} {2013})}\BibitemShut {NoStop}%
\end{thebibliography}%

\end{document}